\documentclass[aps,prb,superscriptaddress,twocolumn,floatfix]{revtex4-1}
\usepackage{graphicx}  
\usepackage{dcolumn}   
\usepackage{bm}        
\usepackage{amssymb}   
\usepackage{color}
\usepackage{pdfpages}
\usepackage{amsmath}   
\usepackage{comment}
\begin{document}

\title{Topological edge plasmon modes between diatomic chains of plasmonic nanoparticles}

\author{C. W. Ling}\affiliation{Department of Applied Physics, The Hong Kong Polytechnic University, Hong Kong, China}
\author{Meng Xiao}\affiliation{Department of Physics, The Hong Kong University of Science and Technology, Hong Kong, China}
\author{C. T. Chan}\affiliation{Department of Physics, The Hong Kong University of Science and Technology, Hong Kong, China}
\author{S. F. Yu}\affiliation{Department of Applied Physics, The Hong Kong Polytechnic University, Hong Kong, China}
\author{Kin Hung Fung}\email{khfung@polyu.edu.hk}\affiliation{Department of Applied Physics, The Hong Kong Polytechnic University, Hong Kong, China}
\date{December 9, 2014}


\begin{abstract}
 We study the topological edge plasmon modes between two ``diatomic'' chains of identical plasmonic nanoparticles. Zak phase for longitudinal plasmon modes in each chain is calculated analytically by solutions of macroscopic Maxwell's equations for particles in quasi-static dipole approximation. This approximation provides a direct analogy with the Su-Schrieffer-Heeger model such that the eigenvalue is mapped to the frequency dependent inverse-polarizability of the nanoparticles. The edge state frequency is found to be the same as the single-particle resonance frequency, which is insensitive to the separation distances within a unit cell. Finally, full electrodynamic simulations with realistic parameters suggest that the edge plasmon mode can be realized through near-field optical spectroscopy.
\end{abstract}


\maketitle
\section{Introduction}
Topology band theory has explained many important electronic phenomena in condensed matter physics, like quantum hall effect, topological insulators, and edge states of 2D graphenes \cite{topologicalinvariant,topologicalorigin,topologicalqhe,symmetrymeets,berryphaseeffect}. Apart from electronic systems, the theory has recently enriched the physics of some classical photonic systems \cite{tphotonreview,phchan,alexander,zheng,YANNOPAPAS1,YANNOPAPAS2}. It turns out that classical topological modes can also be supported in plasmonic systems in the nanoscale \cite{,YANNOPAPAS}. Recently, it is found that topological Majorana-like plasmon modes can be supported in a single chain of plasmonic nanoparticles \cite{aaay} even though the free-space environment is a pass band for photons \cite{radtopo}. However, the toy model in the eigenvalue problem of Ref.~\cite{aaay} employed the pole approximation with several fitting parameters.

One-dimensional chains of plasmonic nanoparticles have been intensively studied in recent years because of their abilities of guiding and confining light in the nanoscale \cite{SPChainEnergya,SPChainEnergyb,SPChainEnergyc,SPChainEnergyd} using less materials as compared to photonic crystals. The guided modes in chains of plasmonic nanoparticles are due to the coupling of localized plasmons among plasmonic nanoparticles. In that sense, each plasmonic nanoparticle plays a role of an ``atom" supporting nearly localized ``orbitals" for photons. If we consider a dimer made of two plasmonic nanoparticles as a unit cell, an array of these dimers can be considered as a ``crystal" domain (simply called a ``diatomic" chain). Here, we study the topologically protected localized plasmon modes formed between the two domains. The Zak phase and the edge mode frequency in ``diatomic" plasmon chains are studied analytically with the Drude material model and the quasi-static dipole approximation for spherical nanoparticles, which is based on the macroscopic Maxwell's equations and is widely accepted. We show that this approximation in classical electrodynamics provides a direct analogy with the Su-Schrieffer-Heeger (SSH) model \cite{SSH,SSHgraphene} for electrons except that the eigenvalue is mapped to the inverse-polarizability of individual nanoparticles. Without any pole approximation of the polarizability, we find that those localized plasmonic modes are topologically protected and the mode frequency is insensitive to the separation distances within a unit cell. The edge state properties are further studied by full electrodynamic simulations of the plasmon enhanced photon emission rate using realistic parameters, which suggests a feasible way to realize the topological plasmon mode\cite{DiracLikePlasmonHoneycomb} through near-field optical spectroscopy.

The paper is organized as follows. We first describe our eigenvalue problem for the plasmon modes and the mapping of the eigenvalues to the mode frequencies in Section \ref{se:EigProblem}. Then, a brief description of the calculation and interpretation of Zak phase for the plasmonic ``diatomic'' chain is given in Section \ref{se:Zak}. The numerical and analytical solutions for the plasmonic edge mode are provided in Sections \ref{se:Solution}. Results of full dynamic simulations are discussed in Section \ref{se:Emission}.
\section{Plasmonic diatomic chain}
We start by calculating the plasmon band dispersion of an infinite ``diatomic'' chain of identical plasmonic nanoparticles (see Figure \ref{fig:geometry1}). As long as the nanoparticles are not too close together, we can model the electromagnetic response of the $n$-th nanoparticle by an electric dipole moment $\bold{p}_n$. The dipole moments satisfy the self-consistent equation ~\cite{weber,coupled}
\begin{equation} \label{eq:couple0}
\alpha^{-1}\bold{p}_n=\sum\limits_{m \neq n} \bold{G}_{nm}\bold{p}_m,
\end{equation}
where $\alpha$ is the polarizability of one nanoparticle and $\bold{G}_{nm}$ is the dyadic Green's function~\cite{jackson} representing the coupling between the $n$th and the $m$th dipole moment. Although Eq. (\ref{eq:couple0}) only forms a set of self-consistent equations for classical electromagnetic waves without showing the Hamiltonian, we can construct a close analogy with known electronic models. Here, we first ignore the long-range couplings and retardation effect and only consider the nearest-neighbor coupling. There are transverse ($xy$ component) and longitudinal mode ($z$ component) in this one-dimensional chain system. It has been shown that the nearest neighbor approximation is pretty good in the longitudinal case \cite{weber} because it is orthogonal to free transverse photon mode propagating in the same direction. For the longitudinal modes, the coupled dipole equation is reduced to:
\begin{equation} \label{eq:couple}
    {\alpha}^{-1}{p_n} =
    \left\{
        \begin{array}{ll}
         \frac{{ 2}}{{4\pi {\epsilon _0}}}
           \left[
                \frac{{ {p_{n - 1}} }}{{ {\left( {{d} - {t}} \right)}^3 }}
                + \frac{{ p_{n + 1} }}{{ {t}^3 }}
           \right],
           & \mbox{for $n$ is even}\\

         \frac{{ 2}}{{4\pi {\epsilon _0}}}
           \left[
                \frac{{p_{n - 1}}}{{{{t}^3}}}
                + \frac{{p_{n + 1}}}{{ {\left( {{d} - {t}}\right)}^3 }}
           \right],
           & \mbox{for $n$ is odd}\\
        \end{array}
     \right.
,\end{equation}
where $\epsilon_0$ is the permitivity of free space, $t$ and $d$ are geometrical parameters defined in Fig.~\ref{fig:geometry1}. It should be noted that all nanoparticles have the same electric dipole polarizability $\alpha(\omega)$, which depends on the angular frequency $\omega$.

\subsection{Eigenvalue problem} \label{se:EigProblem}
Applying the Bloch's theorem to the system, the dipole moments can be written in the form
\begin{equation}\label{eq:pn}
p_n (k) =
    \left\{
        \begin{array}{ll}
        p_A (k)
        e^{i k \frac{{n}}{{2}} d },
        & \mbox{for $n$ is even}\\
        p_B (k)
        e^{i k \left(\frac{{n-1}}{{2}} d \right)},
        & \mbox{for $n$ is odd} \\
        \end{array}
    \right.,
\end{equation}
where $k$ is the wave number of the guided plasmon mode. Substituting the above into the coupled dipole equations Eq.~(\ref{eq:couple}), we obtain a 2 level eigenvalue problem, in which $\alpha^{-1}$ is treated as the eigenvalue:
\begin{equation}\label{eq:2level}
    \begin{pmatrix}
        0       &   a_{12}(k)  \\
        a_{21}(k)  &   0       \\
    \end{pmatrix}
    \begin{pmatrix}
    p_A \\
    p_B \\
     \end{pmatrix}
    =
        \alpha^{-1}
    \begin{pmatrix}
        p_A \\
        p_B \\
    \end{pmatrix},
\end{equation}
where $
a_{12}(k)
= 2 /\left(4 \pi \epsilon_0 \right)
 \{1/ {t}^3+ e^{ - i k d}/{\left(d-t\right)}^3\}
$ and $
a_{21}(k)
= 2 /\left(4 \pi \epsilon_0 \right)
 \{1/ {t}^3+ e^{ i k d}/{\left(d-t\right)}^3\}
$. Although the $2\times2$ matrix in Eq.~(\ref{eq:2level}) is not a Hamiltonian, Eq.~(\ref{eq:2level}) is a Hermitian eigenvalue problem that is analogous to the eigenvalue problem in the SSH model. In this case, the inverse polarizability $\alpha^{-1}$ acts like the energy eigenvalue. There is a mapping between the eigenvalue $\alpha^{-1}$ and the plasmon frequency $\omega$, depending on the materials of the nanoparticles and the range of interest. Using the Drude model \cite{weber,SPChaintheory} for the dielectric function of the materials $\epsilon(\omega)/\epsilon_0=1-{\omega_p}^2/{(\omega^2+i\omega/\tau)}$ and the quasi-static particle polarizability \cite{StaticPolarizability} $\alpha(\omega)=4\pi\epsilon_0 a^3[\epsilon(\omega)-\epsilon_0]/[\epsilon(\omega)+2\epsilon_0]$, the mapping is then
\begin{equation}\label{eq:mapping}
\alpha^{-1}(\omega)=\frac{\frac{1}{3}-\frac{\omega^2}{\omega_p^2}-i(\frac{1}{\tau})(\frac{\omega}{\omega_p^2})}{V\epsilon_0}.
\end{equation}
Here, $\omega_p$, $\tau$, and $V=4\pi a^3/3$ are the plasma frequency, electron mean free time, and volume of the sphere, respectively. It should be noted that $1/\tau$ is the damping coefficient which equals to the electron collision frequency. The above mapping is also plotted in Fig.~\ref{fig:inversealpha} for clear illustration. With the above mapping between the eigenvalue $\alpha^{-1}$ and frequency $\omega$, the non-trivial solutions of the eigenvalue problem in Eq. (\ref{eq:2level}) gives the dispersion relation
\begin{equation}\label{eq:drude}
\alpha^{-1}(\omega)=\pm \sqrt{a_{12}(k)a_{21}(k)}.
\end{equation}

\begin{figure}[htbp]
\centering
\includegraphics[width=3.2in]{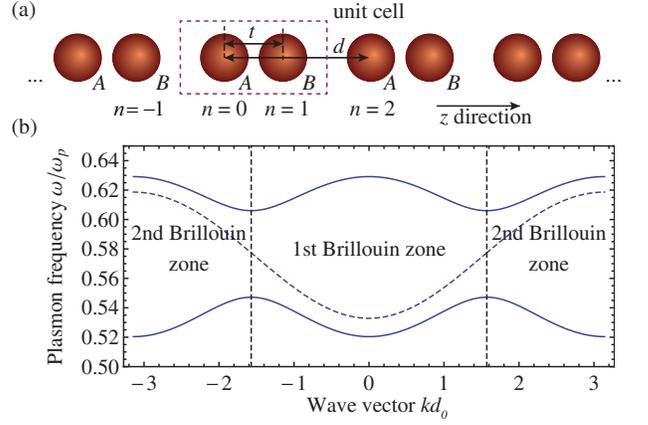}
\caption{\small Band dispersion of the longitudinal dipolar modes in a ``diatomic'' chain of plasmonic nanoparticles. (a) Schematic of the chain aligned in the $z$-direction. There are two spherical metal nanoparticles in a unit cell, and are denoted by sphere $A$ and $B$. The length of unit cell, the separation between spheres (within a unit cell), and the radii of the spheres are denoted by $d$, $t$, and $a$ respectively. The chain is embedded in air. We use $d_0 \equiv d/2$ as the reference length. This figure is drawn in scale such that $a=0.33d/2$, and $t=0.8d/2$. (b) shows the longitudinal mode dispersion relation of the diatomic chain. There are two non-degenerated longitudinal bands (solid lines) as there are two atoms in a unit cell. The dispersion of a monatomic chain is also plotted (dashed line) for comparison. In that case, $t=d/2$, which means the spheres are equally separated.}
\label{fig:geometry1}
\end{figure}

If we neglect the absorption by putting $1/\tau =0$, then we have two real dispersive bands for the system (See Fig.~\ref{fig:geometry1}). The $-$ and $+$ sign in Eq.~(\ref{eq:drude}) refers to the lower and the upper band in the figure, respectively. We see that there are two plasmon bands separated by a gap at $0.606\omega_p>\omega>0.547\omega_p$, which is caused by the two types of coupling between adjacent spheres \cite{interaction}.

The advantage of reformulating the coupled dipole equation in the form of Eq.~(\ref{eq:2level}) is that the equation is still valid when there is material absorption. The matrix is still Hermitian even when we have non-zero damping coefficient. In this situation, the eigenvalue $\alpha^{-1}$ is real, but the quasi-normal modes are having complex plasmon frequency $\omega$. The real and imaginary part of the mode frequency are given by
\begin{equation}\label{eq:imaginary}
    \begin{array}{ll}
    $Re$(\omega)=\sqrt{\frac{{\omega_p}^2}{3}-\frac{1}{4\tau^2}\mp V\epsilon_0{\omega_p}^2\sqrt{a_{12}(k)a_{21}(k)}},\\
    $Im$(\omega) = -\frac{1}{2\tau},\\
    \end{array}
\end{equation}
from which we see the real part will be lowered when damping coefficient $1/\tau$ increases slightly, in the case of small damping.
\begin{figure}[htbp]
\centering
\includegraphics[width=3.2in]{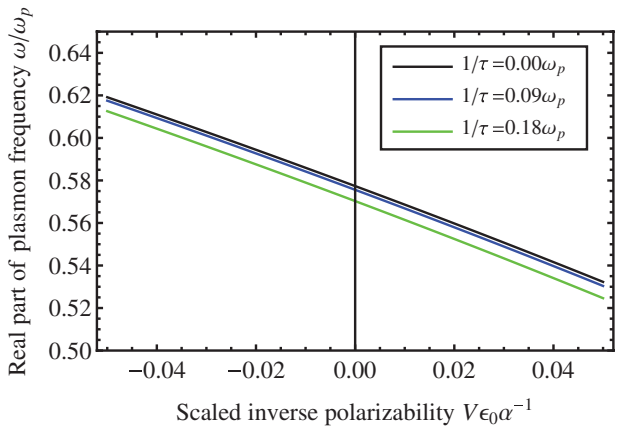}
\caption{\small Mapping between plasmon frequency $\omega$ and the eigenvalue $\alpha^{-1}$ for Drude model of different damping coefficient $1/\tau$ (see Eq.~\ref{eq:mapping}). The mapping is nearly linear in the range of interested shown in Fig.~\ref{fig:geometry1} (b) (i.e., $0.64\omega_p>\omega>0.50\omega_p$). The real part decreases when one increases the damping coefficient.}
\label{fig:inversealpha}
\end{figure}

\subsection{Calculation of Zak phase} \label{se:Zak}

\begin{figure}[htbp]
\centering
\includegraphics[width=3.2in]{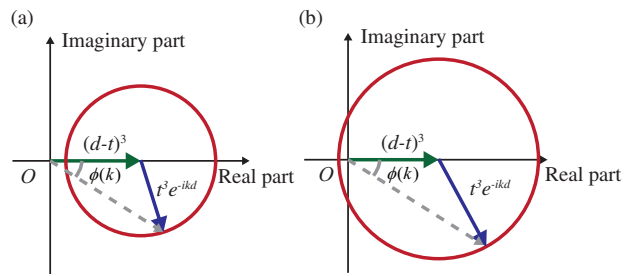}
\caption{\small Representation of $(d-t)^3+{t^3}{\exp(-ikd)}$ in complex plane. The complex number evolves as $kd$ changes from $-\pi$ to $\pi$ for (a) $t < d/2$ and (b) $t > d/2$. In (b), winding number is non-zero, which leads to non-zero Zak phase.}
\label{fig:winding}
\end{figure}

As an analogy of the Berry phase \cite{berryphaseeffect}, the Zak phase \cite{zakphase} has been used to classify band topology for studying their edge states \cite{zakssh,SSH}. We now evaluate the Zak phase using the nontrivial solutions of Eq.~(\ref{eq:2level}). Provided that $t \ne d/2 $, the solution is
\begin{equation}\label{eq:papb}
    \left({
        \begin{array}{*{20}{c}}
           {{p_A (k)}}  \\
           {{p_B (k)}}  \\
        \end{array}
    }\right)
    = \frac{1}{{\sqrt 2 }}
    \left( {
        \begin{array}{*{20}{c}}
            { \pm {e^{i\phi \left( k \right)}}}  \\
            1  \\
        \end{array}
    }\right),
\end{equation}
with $\phi (k) = \arg [ {{{\left( {{d} - {t}} \right)}^3} + {t}^3{e^{ - ik{d}}}} ]$. The above solution is not unique in general (up to a phase factor), which is known as gauge freedom. Since the matrix equation Eq.~(\ref{eq:2level}) is periodic in $k$, a natural choice is requiring that $p_n(k) = p_n (k+G)$, where $G=2 \pi /d$ is the reciprocal lattice vector, and is known as a choice of periodic gauge \cite{periodicgauge}. Equation~(\ref{eq:papb}) satisfies periodic gauge already, and this gauge leads to a $Z_2$ invariant in calculating the Zak phase \cite{zakphonon}:
\begin{equation}\label{eq:zak}
\begin{array}{lll}
 \gamma
    &= i\int_{ {- \frac{\pi}{d}}}^{\frac{\pi}{d}} {\left( {p_A}^*\frac{\partial p_A}{\partial k} + {p_B}^*\frac{\partial p_B}{\partial k} \right)dk}  \\
    & = -\frac{{\phi \left(\frac{\pi }{d}\right) - \phi \left( - \frac{\pi }{d}\right)}}{2} \\
 \end{array}
 \end{equation}
For example, if $t < d/2$, we have $\phi(\pi/d)-\phi(-\pi/d)=0$. On the other hand, if $t > d/2$, then $\phi(\pi/d)-\phi(-\pi/d)=-2\pi$, see Fig.~\ref{fig:winding}.
Here, we have the same $\gamma$ for both the lower and the upper bands because the $\pm$ sign in Eq.~(\ref{eq:papb}) for different bands are canceled when multiplying by its conjugate. The Zak phase can be interpreted using the winding number in this plasmonic system, as similar to those in graphene \cite{SSHgraphene}, see Fig.~\ref{fig:winding}. From this we can classify the system into two classes, one with $\gamma=\pi$ and one with $\gamma=0$, see Table ~\ref{tab:zak}. It should be noted that the choice of unit cell would also lead to different Zak phases. For example, if we shift the unit cell by half the period, the separation between atoms within the unit cell will change from $t$ to $d-t$, and $\gamma$ will change by $\pi$. The choice of unit cell is arbitrary for an infinite chain but not a truncated chain where the boundary conditions limit the choice.
\begin{table}[h]{
\caption{Zak phase $\gamma$ of diatomic chains. The upper and the lower bands share the same value of Zak phase.}\label{tab:zak}
\begin{tabular}{l | c | c}

                    & \multicolumn{2}{c}{sphere $A$ and $B$ as atoms in unit cell}\\
                    \cline{2-3}
                    & \parbox[c]{2.2 cm}{upper band}    & \parbox[c]{2.2 cm}{lower band} \\ \hline
   $\gamma$ for $t < d/2$    &  0                       &  0           \\ \hline
   $\gamma$ for $t > d/2$    & $\pi$                    &  $\pi$        \\ \hline
\end{tabular}}
\end{table}

\section{Plasmonic edge mode in connected chain} \label{se:Solution}
The Zak phase gives a simple topological classification of the plasmon bands. If two chains with different Zak phases are connected together to form a new chain, it is expected that there exists an edge state localized at the interface between the chains. To verify the existence of such an edge state, we now consider two semi-infinite diatomic chains connected together [see Fig.~\ref{fig:edgestate} (a)]. The left and the right chains have the same lattice constant $d$, and the separations between the two atoms in the unit cell are $t=t_L$ and $t=t_R$ respectively.

\subsection{Numerical solution for the edge mode}
\begin{figure}[htbp]
\centering
\includegraphics[width=3.2in]{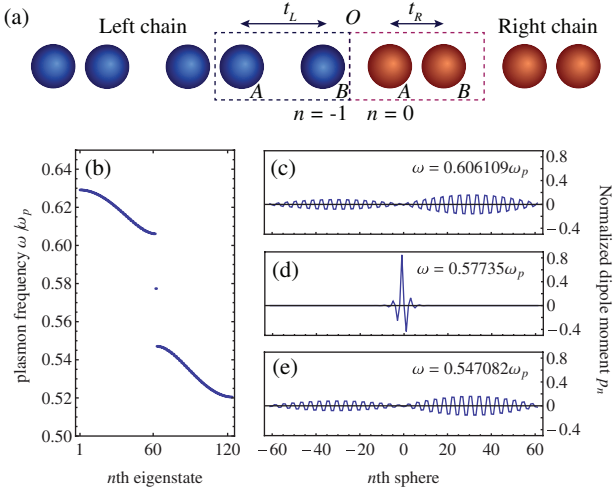}
\caption{\small Edge states between diatomic chains. (a) shows the geometry of a connected chain system. Separation $t=t_L$ for the left chain and $t=t_R$ for the right chain. The interface is in between $n=-1$ and $n=0$. (b) shows the plasmon frequency obtained by solving eigenvalue problem of a connected chain with 123 spheres. $t_L$, $t_R$, and $a$ are set to $1.2 d/2$, $0.8 d/2$, and $0.33 d/2$. (a) is drawn in scale with these parameters. (c) and (e) show the eigenstate of the connected chain just on top and below the band gap. (d) shows the edge state at the interface.}
\label{fig:edgestate}
\end{figure}

Here, we first consider a finite chain connected by two ``diatomic'' chains with $t_L=1.2 d/2$ and $t_R=0.8 d/2$. The left chain is from $n=-61$ to $n=-1$, and the right chain is from $n=0$ to $n=61$. By adopting the lossless Drude model Eq.~(\ref{eq:mapping}), we numerically solve the finite coupled dipole equation Eq.~(\ref{eq:couple}) by rewriting $\omega^2/{\omega_p}^2$ as eigenvalue. The plasmon frequencies of the eigenstates are shown in Fig.~\ref{fig:edgestate}(b). The results confirm that there is a band gap in the region $0.606\omega_p>\omega>0.547\omega_p$. The figure also indicates a state within the band gap, with $\omega=0.577\omega_p$. We show the dipole moments $p_n$ of the system for three states in Figs.~\ref{fig:edgestate} (c) to (e). Figure ~\ref{fig:edgestate} (c) and (e) show the states just above and below the band gap while Fig.~\ref{fig:edgestate} (d) shows the edge state at the interface of chains. The magnitudes of dipole moments in Fig.~\ref{fig:edgestate} (d) decay away from the interface. The edge state is neither symmetric nor antisymmetric about the interface because the system has a broken inversion (and reflection) symmetry. We also verified numerically that there is no edge state in between the chains when both the left and right chains are sharing the same Zak phase.

\begin{figure}[htbp]
\centering
\includegraphics[width= 2.9 in]{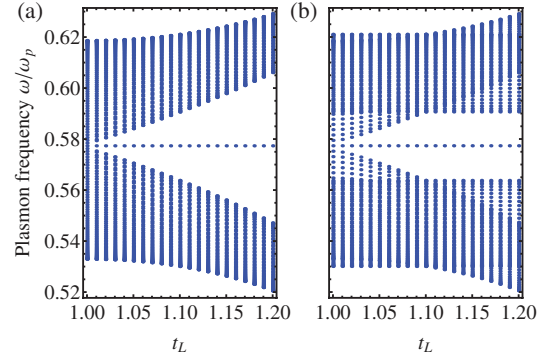}
\caption{\small Dependence of bands and edge state on $t_L$.  (a) $t_R$ is set equal to $d-t_L$. For example when $t_L=1.1 d/2$, $t_R=0.9 d/2$. (b) $t_R$ is fixed and equals $0.9 d/2$. The frequency of edge state does not vary with $t_L$ in both cases.}
\label{fig:frq_vs_tl}
\end{figure}

Considering the same eigenvalue problem, we can show the robustness of the edge state. The plasmon frequencies of the system (123 spheres) with different $t_L$ and $t_R$ are shown in Fig.~\ref{fig:frq_vs_tl}. We first set $t_R=d-t_L$ [See Fig.~\ref{fig:frq_vs_tl} (a)]. In this case, we see a single edge state at the interface when $t_L\neq d/2$. It is very interesting to note that the state frequency is independent of $t_L$. In Fig.~\ref{fig:frq_vs_tl} (b), we fix $t_R=0.9 d/2$ but varies $t_L$. Similar to the case in Fig.~\ref{fig:frq_vs_tl} (a), a single edge state exists when $t_L\neq d/2$ and the edge state frequency is independent of the variable parameters (although the band frequencies change a lot).

\subsection{Analytical solution for the edge mode}
In the following, with $1/\tau=0$, we show explicitly that there is always an edge state solution when the two semi-infinite chains have different Zak phases, that is $t_R<d/2<t_L$ or $t_R>d/2>t_L$. The state is localized at the boundary between the two chains and always has frequency $\omega/\omega_p=1/\sqrt{3}$ (Fig.~\ref{fig:frq_vs_tl}). Consider the case $t_R<d/2<t_L$ as shown in Fig.~\ref{fig:edgestate}(a).  Dividing the connected chain into two semi-infinite regions, the right and the left chain, the plasmon wave vector $k$ takes different forms, $k= \pi/d+i\mu_R$ and $k=-\pi/d-i\mu_L$, in the two regions. Here $\mu_R$ and $\mu_L$ are real positive constants, and are related to decay length in different regions \cite{BasicSurface}. Substituting it into $a_{12}$ and $a_{21}$, and using superscript $^R$ and $^L$ to denote the expressions for the right and the left chain, we have
\begin{equation*}
    \begin{array}{ll}
    a_{12,21}^{R}=\frac{{2}}{{4\pi\epsilon_0}}\left(\frac{{1}}{{t_{R}^3}}-\frac{{1}}{{{(d-t_R)}^3}}e^{\pm \mu_{R}d}\right)  \\
    a_{12,21}^{L}=\frac{{2}}{{4\pi\epsilon_0}}\left(\frac{{1}}{{t_{L}^3}}-\frac{{1}}{{{(d-t_L)}^3}}e^{\mp \mu_{L}d}\right)  \\
    \end{array}
\end{equation*}
In the above, $a_{12}$ corresponds to upper sign while $a_{21}$ corresponds to lower sign. It can be inferred from Eq.~(\ref{eq:2level}) that non-trivial solution exists when
\begin{equation}\label{eq:alphainverse}
(\alpha^{-1})^2=a_{12}^R a_{21}^R=a_{12}^L a_{21}^L.
\end{equation}
For $\omega/\omega_p=1/\sqrt{3}$, Eq.~(\ref{eq:drude}) implies that $\alpha^{-1}=0$ (i.e., at least one of $a_{12}^R$ and $a_{21}^R$ is zero and at least one of $a_{12}^L$ and $a_{21}^L$ is zero). When $t_L<d/2$, the factor $1/t_R^3-e^{(-\mu_R d)}/(d-t_R)^3$ in $a_{21}^R$ will never equal zero as $\mu_R$ varies. Therefore, in this case, $a_{21}^R\neq0$ and $a_{12}^R=0$, which implies $e^{\mu_R d} = (d-t_R)^3/(t_R)^3$. Similarly,  we have $a_{21}^L\neq0$, $a_{12}^L=0$ and $e^{-\mu_L d} = (d-t_L)^3/(t_L)^3$ . Using Eq.~(\ref{eq:2level}) with $a_{12}^R=0$ and $a_{12}^L=0$, the corresponding solutions in the two regions are
\begin{equation}
    \begin{pmatrix}
    p_A^R\\
    p_B^R\\
    \end{pmatrix}
    =
    \begin{pmatrix}
    0\\
    1\\
    \end{pmatrix}
    \mbox{and}
    \begin{pmatrix}
    p_A^L\\
    p_B^L\\
    \end{pmatrix}
    =
    \begin{pmatrix}
    0\\
    1\\
    \end{pmatrix}.
\end{equation}
With ${p_n}^R$ and ${p_n}^L$ denoting, respectively, the solution for the right and the left region, the edge state solution is written in the form
\begin{equation}\label{eq:pnconnected}
{p_n} = \left\{
 {\begin{array}{ll}
   C{p_n}^R, & \mbox{for $n \geqslant 0$ (right region)}  \\
   D{p_n}^L, & \mbox{for $n < 0$ (left region)}  \\
\end{array} } \right.,
\end{equation}
where $C$ and $D$ are constants to be determined by matching with the interface equations:
\begin{equation}\label{eq:boundary2}
\left\{
    \begin{array}{ll}
    4\pi\epsilon_0\alpha^{-1}p_{-1}=\frac{{2}}{{(d-t_0)^3}}p_0+\frac{{2}}{{t_L^3}}p_{-2}\\
    4\pi\epsilon_0\alpha^{-1}p_0=\frac{{2}}{{(d-t_0)^3}}p_{-1}+\frac{{2}}{{t_R^3}}p_1 \\
    \end{array},
\right.
\end{equation}
in which $t_0 = (t_L+t_R)/2$. With Eq.~(\ref{eq:pn}) and (\ref{eq:pnconnected}), we have dipole moments near the interface $(p_{-2},p_{-1},p_{0},p_{1})=(0,D,0,C)$. Putting into Eq.~(\ref{eq:boundary2}), we have $C/D=-{t_R}^3/(d-\frac{{t_L+t_R}}{{2}})^3$. Thus, the edge state solution is
\begin{equation}\label{eq:edgestate}
p_n=
\left\{
    \begin{array}{ll}
    0,                                                  &\mbox{for $n$ is even}             \\
    (-1)^{\frac{{n+1}}{{2}}}e^{\frac{{n+1}}{{2}}\mu_L d},  &\mbox{for $n$ is odd and $n<0$}    \\
    (-1)^{\frac{{n+1}}{{2}}}\frac{{t_R^3}}{{(d-(t_L+t_R)/2)^3}}\\
     \times e^{-\frac{{n-1}}{{2}}\mu_R d},    &\mbox{for $n$ is odd and $n>0$}   \\
    \end{array}
\right.
\end{equation}
This shows that the edge state exist at a frequency of $\omega/\omega_p=1/\sqrt{3}=0.577$. The presence of exponential factors in the edge state indicate dipole moments decay away from the interface.

The edge state in Fig.~\ref{fig:edgestate}(d) can be confirmed by putting $t_R = d- t_L$ in Eq.~(\ref{eq:edgestate}). In this case, we further show that there is no edge state solution within band gap when $\omega/\omega_p \neq 1/ \sqrt{3}$. By expanding expressions in Eq.~(\ref{eq:alphainverse}), one have $\mu_R=\mu_L\equiv\mu$ and $\alpha^{-1} \neq 0$, so the non-trivial state for the two level problem is
\begin{equation}
    \begin{array}{lll}
        \begin{pmatrix}
          p_A^R \\
          p_B^R \\
        \end{pmatrix}
        =
        \begin{pmatrix}
          a_{12}^R/\alpha^{-1} \\
          1 \\
        \end{pmatrix}
        &
        \mbox{and}
        &
        \begin{pmatrix}
          p_A^L \\
          p_B^L \\
        \end{pmatrix}
        =
        \begin{pmatrix}
          a_{12}^L/\alpha^{-1} \\
          1 \\
        \end{pmatrix}
    \end {array}.
\end{equation}
Normalizing factor in the above is absorbed to the tuning constants $C$ and $D$ in the next step. Substituting the above into Eq.~(\ref{eq:edgestate}), we have dipole moments near the interface
$(p_{-2},p_{-1})=D\left(a_{12}^L/\alpha^{-1},1\right)$ and $(p_0,p_1)=C\left(a_{12}^R/\alpha^{-1},1\right)$
. Putting $p_{-2}$, $p_{-1}$, $p_0$, and $p_1$ into Eq.~(\ref{eq:boundary2}), one can verify that $C=D=0$, which means there is no connected chain state whose frequency lies inside the band gap but $\omega/\omega_p \neq 1/\sqrt{3}$.

\section{Emission rate} \label{se:Emission}
So far, our calculations are based on quasi-static approximations. Here, we verify the existence of the edge mode calculated in section \ref{se:Solution} by full-wave simulation using the finite-difference time-domain method. Lumerical FDTD, a commercial-grade simulator, was used to perform the calculations\cite{FDTD}. To do this, we put a dipole emitter inside one of the nanoparticles. Using mesh size $=1$ nm, we simulated the connected chain with parameters $a=25$ nm, $d=150$ nm, $t_L=90$ nm, and $t_R=60$ nm (so $t_L=1.2 d/2$ and $t_R=0.8 d/2$). Material permittivity is a Drude model with plasma frequency $\omega_p=1\times10^{16}$ ${\rm{rads^{-1}}}$ and electron collision frequency $1/\tau=3\times10^{14}$ ${\rm{rads^{-1}}}$. The left chain runs from sphere $n=- 13$ to $n=-1$, while the right chain runs from sphere $n=0$ to $n=11$. The dipole source is put inside the sphere $n=-1$, as it will attain maximum dipole moment in the edge state, and so we can have prominent results. Also, we etched a hole with radius 6 nm in the sphere $n=-1$ to avoid contact between the dipole source and the metallic structures, which eventually leads to diverging numerical errors.

\begin{figure}[htbp]
\centering
\includegraphics[width= 2.7 in]{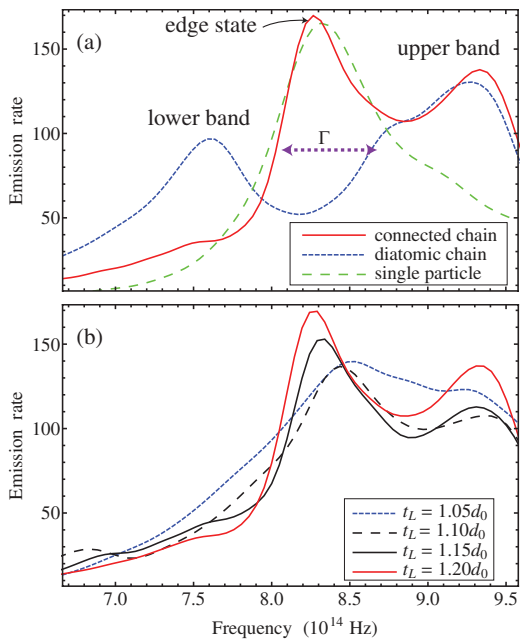}
\caption{\small Photon emission rates in plasmonic nanoparticle chains calculated by finite-difference time-domain simulations. A dipole source is placed inside a particular sphere with a small hole of negligible size, acting as an emitter. The emission rate is defined by Eq.~(\ref{eq:emission}), which represents approximately the local density of states (LDOS). (a) Emission rate in single sphere, diatomic chain, and connected chain. The last one reveals the existence of an edge state. Videos of corresponding time-domain fields are attached, in which longitudinal component of electric fields are shown. \textcolor{blue}{Media~1}: connected chain; \textcolor{blue}{Media~2}: diatomic chain; \textcolor{blue}{Media~3}: monatomic chain; \textcolor{blue}{Media~4}: single sphere . (b) Emission rate for connected case with different $t_L$. $d/2 = 75$ nm and $t_R = d-t_L$.}
\label{fig:emittance}
\end{figure}

We define an emission rate, which has a meaning similar to the local density of states (LDOS), as the total power flowing out of the metal nanoparticle divided by the source power:
\begin{equation}\label{eq:emission}
{\text{emission rate}} (\omega) = \frac{{ {\text{power flowing out}}(\omega) }}{{ {\text{source power}} (\omega)}},
\end{equation}
where $\omega$ is the dipole source angular frequency. Here, the source power means the dipole emission power in free space. The presence of a plasmon mode would result in larger emission rate, which can help us to locate the response and edge states. The rate is found by boxing the whole sphere containing the dipole source with power monitors. Let us first focusing on the emission rate for the connected chain (contains 25 spheres) as shown in Fig.~\ref{fig:emittance} (a). We see that there are two ``prominent hills" peaked at 8.3 and $9.4\times10^{14}$ Hz, and one ``small hill" peaked at $7.6\times10^{14}$ Hz. The middle hill within the band gap is due to the presence of the edge state, and the right and the left hills are the responses of the upper and the lower band. A comparison with a pure diatomic chain (where we set both $t_L=t_R= 60$ nm, contains 24 spheres) is also shown in Fig.~\ref{fig:emittance}. In this case, there are two prominent hills peaked at about 7.6 and $9.4\times10^{14}$ Hz, which are corresponding to the upper and the lower band. The valley in between, is noticed as a band gap.

It should be noted that the emission rate depends on the source position. For the case of the connected chain in Fig.~\ref{fig:emittance} (a), the small hill at $7.6\times10^{14}$ Hz will become prominent if we put the dipole source and measure the emission rate of sphere $n=-2$ instead. The edge state peak coincides with the the single sphere resonance frequency $8.3\times10^{14}$ Hz, which is consistent to the result $\omega/\omega_p=1/\sqrt{3}$. In fact there is one more peak at a higher frequency for the single sphere, as there are two metal-dielectric surfaces\cite{shell}, but it is outside the frequency range in the figure and so it is not shown here.

In Fig.~\ref{fig:emittance} (b), we showed the emission rate for connected chain with different $t_L = 1.05 d/2$, $1.10 d/2$, $1.15 d/2$, and $1.20 d/2$, where $d/2=75$ nm. Each curve contains an edge state peak around $8.3\times10^{14}$ Hz, which shows the invariance of the edge state frequency. For the case $t_L=1.05 d/2$, as it is close to a monatomic plasmon chain, the band gap is no longer visible.

The presence of damping leads to broadening of resonant peaks. Here, we would like to show that the full width half maximum(FWHM) $\Gamma$ of the edge state in the photon emission rate can be roughly estimated by considering damping the material model. Since the dipole wave is in the form of $p_n(k)\exp{[-{\text{Im}}(\omega)t]}\exp{[i{\text{Re}}(\omega)t]}$, the emission peak thus has a width of $\Gamma=2{\text{Im}}(\omega)$ \cite{jackson}. Together with Eq.~(\ref{eq:imaginary}), we can estimate the width as $\Gamma=1/\tau = 3\times 10^{14}$ rad/s. In numerical simulation, it should be noted that the edge state resonant peak and the upper band can have overlap [see Fig.~\ref{fig:emittance}(a)], so FWHM in the graph is roughly estimated by reading the trend of edge state peak, which is about $0.7\times 10^{14}$ Hz ($4.4 \times 10^{14}$ rad/s) in length. Although the theoretical value is estimated without consideration of radiation loss, the result is still close to the value obtained from Fig.~\ref{fig:emittance}(a).

\section{Conclusion}
To conclude, we studied the plasmonic topological edge states between diatomic chains of plasmonic nanoparticles by making an analogy with the Su-Schrieffer-Heeger model such that the eigenvalue is mapped to the inverse-polarizability of the nanoparticles. When two diatomic chains with different Zak phases are connected, it is found that a new localized plasmon mode appears at the domain boundary. Even though our analytical results are calculated using quasistatic point dipole approximation, they have correctly predicted the existence of the plasmonic edge modes in full electrodynamic simulation with Drude metal nanoparticles. Our results suggest a feasible way to realize the topological plasmon mode through near-field optical spectroscopy.

\section*{Acknowledgments}
This work was supported by the Hong Kong Research Grant Council through the Early Career Scheme (grant no. 509813) and the Area of Excellence Scheme (grant no. AoE/P-02/12). We thank Z. Q. Zhang, Xueqin Huang, K. T. Law, C. H. Lam, and C. T. Yip for useful discussions.


\end{document}